\documentclass[12pt,english,draftcls,  one column]{IEEEtran}
\usepackage[T1]{fontenc}
\usepackage[latin9]{inputenc}
\usepackage{amsthm}
\usepackage{amsmath}
\usepackage{amssymb}
\usepackage{graphicx}

\makeatletter
\theoremstyle{plain}

\theoremstyle{plain}

\ifCLASSINFOpdf% \usepackage[pdftex]{graphicx}
\else% or other class option (dvipsone, dvipdf, if not using dvips). graphicx
\fi% graphicx was written by David Carlisle and Sebastian Rahtz. It is
\usepackage{babel}

\makeatother

\usepackage{babel}
\providecommand{\propositionname}{Proposition}
\providecommand{\theoremname}{Theorem}

\begin{document}

\title{Multi-Layer Hybrid-ARQ \\for an Out-of-Band Relay Channel}

\author{Seok-Hwan Park, Osvaldo Simeone, Onur Sahin and Shlomo Shamai (Shitz)
\thanks{S.-H. Park and O. Simeone are with the Center for Wireless Communications
and Signal Processing Research (CWCSPR), ECE Department, New Jersey
Institute of Technology (NJIT), Newark, NJ 07102, USA (email: \{seok-hwan.park,
osvaldo.simeone\}@njit.edu).

O. Sahin is with InterDigital Inc., Melville, New York, 11747, USA
(email: Onur.Sahin@interdigital.com).

S. Shamai (Shitz) is with the Department of Electrical Engineering,
Technion, Haifa, 32000, Israel (email: sshlomo@ee.technion.ac.il).%
}}
\maketitle
\begin{abstract}
This paper addresses robust communication on a fading relay channel
in which the relay is connected to the decoder via an out-of-band
digital link of limited capacity. Both the source-to-relay and the
source-to-destination links are subject to fading gains, which are
generally unknown to the encoder prior to transmission. To overcome
this impairment, a hybrid automatic retransmission request (HARQ)
protocol is combined with multi-layer broadcast transmission, thus
allowing for variable-rate decoding. Moreover, motivated by cloud
radio access network applications, the relay operation is limited
to compress-and-forward. The aim is maximizing the throughput performance
as measured by the average number of successfully received bits per
channel use, under either long-term static channel (LTSC) or short-term
static channel (STSC) models. In order to opportunistically leverage
better channel states based on the HARQ feedback from the decoder,
an adaptive compression strategy at the relay is also proposed. Numerical
results confirm the effectiveness of the proposed strategies.

\theoremstyle{theorem}
\newtheorem{theorem}{Theorem}
\theoremstyle{proposition}
\newtheorem{proposition}{Proposition}
\theoremstyle{lemma}
\newtheorem{lemma}{Lemma}
\theoremstyle{corollary}
\newtheorem{corollary}{Corollary}
\theoremstyle{definition}
\newtheorem{definition}{Definition}
\theoremstyle{remark}
\newtheorem{remark}{Remark}
\end{abstract}

\section{Introduction\label{sec:Introduction}}

Consider the fading relay channel model shown in Fig. \ref{fig:system model},
in which an encoder communicates to a decoder through a relay that
is connected to the decoder via an out-of-band capacity-constrained
backhaul link. Both the source-to-relay and the source-to-destination
links are subject to fading. The motivation for this model comes from
the uplink of cloud radio access networks \cite{Huawei}\cite{Marsch},
in which base stations (BSs) operate as soft relays that communicate
with a central decoder via a digital backhaul links. In this scenario,
the central decoder performs decoding based on the compressed signals
collected from the connected BSs. With regards to this application,
in the model of Fig. \ref{fig:system model}, the encoder represents
a mobile station (MS); the relay is the BS in the same cell, which
is connected to the (central) decoder via backhaul link; and the signal
$Y$ represents the compressed signals collected by the decoder from
BSs belonging to other cells. The signal $Y$ can be seen as \textit{side
information} available at the decoder when designing the encoder (i.e.,
the MS) and relay (i.e., the BS).

The fading relay channel was investigated in \cite{Steiner:TwoHop}
and \cite{Khandani} in the absence of the direct link between the
source and the destination and assuming fading also on the relay-to-destination
link. In \cite{Steiner:TwoHop}, various relaying protocols including
decode-and-forward, quantize-and-forward and hybrid amplify-quantize-and-forward
were developed in combination with layered broadcast coding (BC).
This work was extended in \cite{Khandani} by studying infinite-layer
coding at both the source and the relay in conjunction with decode-and-forward
relaying.

The fading relay channels with a direct link between the source and
the destination (as in Fig. \ref{fig:system model}) was studied in
\cite{del Coso}-\cite{delaytol}. The works in \cite{del Coso}-\cite{Tian}
solved the problem of optimizing the compression strategy at the relay
under the assumption of perfect channel state information for multi-antenna
terminals. In the presence of uncertainty on the fading coefficient
$S$, layered approaches that adopt a competitive, rather than average,
optimality criterion are derived in \cite{competitive} and \cite{delaytol}
assuming no hybrid automatic retransmission request (HARQ). In all
the previous works, the feedback link in Fig. \ref{fig:system model}
was not included. This link is used in this paper to enable HARQ.

\subsection{Contributions\label{sub:Contirubitions}}

In this work, motivated by cloud radio access applications as mentioned
above, we study the system in Fig. \ref{fig:system model}, assuming
that the relay performs compress-and-forward. We propose to combine
two key strategies to mitigate the impact of the fading on the source-to-relay
and source-to-destination links, namely, HARQ and BC. With HARQ, the
decoder requests retransmission by sending feedback information to
the encoder and the relay regarding the outcome of the decoding process.
Specifically, the incremental redundancy HARQ (IR-HARQ) consists of
the transmission of additional parity bits in case of failed decoding
\cite{Caire}. With BC \cite{Cover BC}-\cite{Verdu Variable}, instead,
one allows for variable-rate decoding that opportunistically adapts
to the actual fading state conditions.

Multi-layer HARQ strategies have thus the advantage of allowing for
variable-length transmission and variable-rate decoding, and were
introduced in \cite{Steiner:HARQ} for point-to-point fading channels.
As in \cite{Steiner:HARQ}, we aim at maximizing the average throughput
and distinguish two scenarios, namely short-term static channel (STSC)
and long-term static channel (LTSC). Moreover, for the LTSC scenario,
we propose an adaptive compression method at the relay that is able
to opportunistically leverage better fading state based on the feedback
information received from the decoder. The effectiveness of the proposed
multi-layer HARQ strategies is confirmed via extensive numerical results.

The paper is organized as follows. We state the system model in Sec.
\ref{sec:System-Model} and establish the problem formulation in Sec.
\ref{sec:Problem-Definition}. After describing the proposed multi-layer
HARQ strategies with a constant compression gain and adaptive compression
gain in Sec. \ref{sec:fixed compression} and Sec. \ref{sec:Adaptive Compression},
respectively, for the LTSC model, we extend the discussion to the
STSC model in Sec. \ref{sec:STSC model}. Numerical results are provided
in Sec. \ref{sec:Numerical-Results} to demonstrate the performance
gain of the proposed multi-layer HARQ strategies.

\textit{Notation}: We adopt standard information-theoretic definitions
for the mutual information $I(X;Y)$ between the random variables
$X$ and $Y$, and conditional mutual information $I(X;Y|Z)$ between
$X$ and $Y$ conditioned on random variable $Z$ \cite{ElGamal}.
All logarithms are in base two unless specified. We use $\mathbb{E}_{X}[\cdot]$
to denote the expectation over $X$. For a real number $x$, we define
a function $[x]^{+}=\max\{x,0\}$.

\section{System Model\label{sec:System-Model}}

\begin{figure}
\centering\includegraphics[width=14cm,height=5cm]{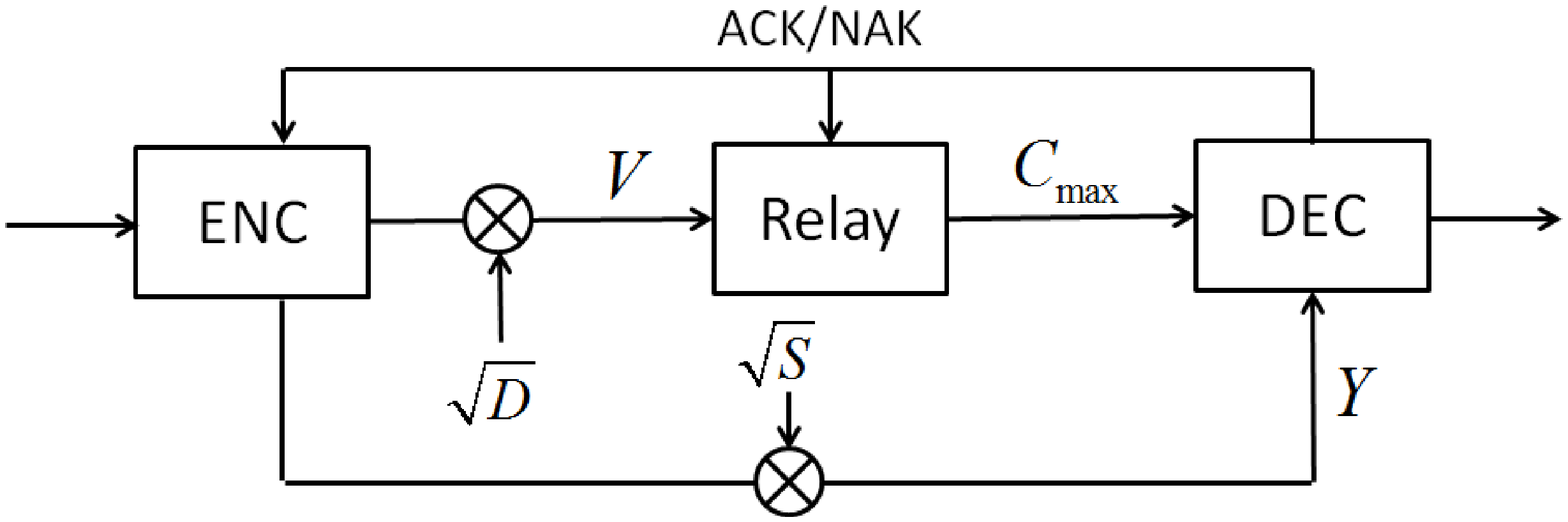}

\caption{\label{fig:system model}A fading relay channel with out-of-band relay-to-decoder
communication and a feedback link to enable HARQ.}
\end{figure}

We consider the fading relay channel depicted in Fig. \ref{fig:system model},
in which the relay is connected to the decoder via a digital link
of capacity $C_{\max}$. In order to enable HARQ, after each transmission
block (or slot), the decoder sends feedback information to the encoder
acknowledging, or not, successful decoding. This feedback link is
assumed to be error-free.

\subsection{Channel Model\label{sub:Channel-Model}}

The signal $V_{t,i}$ received by the relay in the $i$th symbol,
$i=1,\ldots,n$, of the $t$th transmission slot is given as
\begin{equation}
V_{t,i}=\sqrt{D_{t}}X_{t,i}+E_{t,i},\label{eq:received signal local}
\end{equation}
for $t=1,\ldots,T$, where $D_{t}$ is the fading coefficient in the
$t$th time slot, $X_{t,i}\sim\mathcal{N}(0,P)$ represents the signal
transmitted by the encoder, $E_{t,i}\sim\mathcal{N}(0,1)$ is the
additive noise at the relay, and $T$ is the maximum tolerable delay
for the HARQ process. As it will be detailed, upon correct decoding
at the destination, the HARQ process is stopped, and $T$ is the maximum
number of overall transmissions allowed for the same data packet.
We assume that the block size $n$ is large enough to enable the use
of information-theoretic limits. The notation $D_{t}$ has been chosen
with reference to the cloud radio access application in which $D_{t}$
represents the direct channel to the local BS.

The symbol $Y_{t,i}$ received by the decoder in the $i$th symbol,
$i=1,\ldots,n$, of the $t$th transmission slot is
\begin{equation}
Y_{t,i}=\sqrt{S_{t}}X_{t,i}+Z_{t,i},\label{eq:received signal side}
\end{equation}
for $t=1,\ldots,T$, where $S_{t}$ is the fading coefficient in the
$t$th time slot, and $Z_{t,i}\sim\mathcal{N}(0,1)$ is the additive
noise. The notation $S_{t}$ is a reminder that in the cloud radio
access application, $S_{t}$ represents the side information channel
(see Sec. \ref{sec:Introduction}). From now on, we omit the symbol
index $i$ for notational brevity.

Following \cite{Steiner:HARQ}, depending on the channel coherence
time, we distinguish two scenarios: \textit{i)} short-term static
channel (STSC); and \textit{ii)} long-term static channel (LTSC).
With LTSC, the channels remain fixed over all the, at most, $T$ transmission
blocks used for the current data packet, that is,
\begin{equation}
D_{t}=D\,\,\mathrm{and}\,\, S_{t}=S,\,\,\mathrm{for\,\, all}\,\, t=1,\ldots,T.\label{eq:LTSC model}
\end{equation}
In contrast, with STSC, the channel changes independently from block
to block. We first study the LTSC model (\ref{eq:LTSC model}) in
Sec. \ref{sec:fixed compression} and Sec. \ref{sec:Adaptive Compression},
and then consider the STSC case in Sec. \ref{sec:STSC model}. We
assume that the fading coefficients $D_{t}$ in (\ref{eq:received signal local})
and $S_{t}$ in (\ref{eq:received signal side}) are independent,
and have arbitrary CDFs $F_{D}(d)=\Pr[D_{t}\leq d]$ and $F_{S}(s)=\Pr[S_{t}\leq s]$
with finite powers $\rho_{D}=\mathbb{E}[|D|^{2}]$ and $\rho_{S}=\mathbb{E}[|S|^{2}]$,
respectively.

The realization of the fading coefficients $S_{t}$ is known only
to the decoder, while that of the fading coefficients $D_{t}$ is
available at the decoder as well as the relay. In order to study the
effect of the local CSI at the encoder, we will consider both cases
where the encoder knows the realization of the ``direct'' fading
channel $D_{t}$ to the relay, e.g., through feedback, or not.

\subsection{Relay Operation\label{sub:Relay-Operation}}

The relay compresses its received signal $V_{t}$ and sends a description
$W_{t}$ to the decoder. Without claim of optimality, we assume a
Gaussian test channel (see, e.g., \cite{ElGamal}) as
\begin{equation}
W_{t}=\sqrt{a_{t}}V_{t}+N_{t},\label{eq:Gaussian test channel}
\end{equation}
where $a_{t}$ is a non-negative compression gain and $N_{t}\sim\mathcal{N}(0,1)$
represents the compression noise. Using binning for distributed source
coding at the relay by leveraging the side information (\ref{eq:received signal side})
at the decoder, the latter can recover the description $W_{t}$ as
long as the inequality
\begin{equation}
I(V_{t};W_{t}|Y_{t})=\frac{1}{2}\log\left(1+a_{t}\left(\frac{1}{D_{t}}+\frac{P}{1+PS_{t}}\right)\right)\leq C_{\max}\label{eq:backhaul constraint}
\end{equation}
is satisfied \cite[Ch. 11]{ElGamal}. Due to the mentioned CSI limitation,
the relay should compute the compression gain $a_{t}$ as a function
of the realization of the local fading $D_{t}$ without having information
about the fading state $S_{t}$ in (\ref{eq:received signal side}).
Therefore, in order to guarantee that the decoder can always recover
$W_{t}$ regardless of the realization of the fading coefficient $S_{t}$
on the side information (\ref{eq:received signal side}), one needs
to set the compression gain $a_{t}$ so that (\ref{eq:backhaul constraint})
is satisfied even for the minimum value $s_{\min}$ in the support
of $F_{S}(s)$ (i.e., $s_{\min}=\mathrm{inf}\{s:F_{S}(s)>0\}$). This
leads to
\begin{equation}
a_{t}=\frac{\beta(1+s_{\min}P)}{1/d+(1+s_{\min}/d)P}\triangleq a_{d},\label{eq:compression gain conservative}
\end{equation}
where $\beta=2^{2C_{\max}}-1$, for $D_{t}=d$. We will consider different
strategies for the choice of the compression gain $a_{t}$ in Sec.
\ref{sec:fixed compression} and Sec. \ref{sec:Adaptive Compression}.

\subsection{Multi-Layer Hybrid-ARQ\label{sub:Multi-Layer-Hybrid-ARQ}}

Following \cite{Steiner:HARQ}, the encoder uses a two-layer BC transmission
strategy coupled with HARQ, which is described next. The encoder wishes
to deliver two messages $M_{1}\in\{1,\ldots,2^{nR_{1}}\}$ and $M_{2}\in\{1,\ldots,2^{nR_{2}}\}$,
which are independent and uniformly distributed, to the decoder. To
this end, it maps message $M_{l}$ to a $n$-symbol codeword $X_{l,t}^{n}$
for $l=1,2$. We assume independent Gaussian codebooks across the
$T$ blocks, that is, the codewords $X_{l,t}^{n}$ are independently
generated with i.i.d. symbols $\mathcal{N}(0,1)$ for all blocks $t=1,\ldots,T$.
To describe the multi-layer HARQ strategy, we distinguish the following
two transmission modes: \textit{i)} BC mode; and \textit{ii)} single-layer
(SL) mode. In the BC mode, the encoder transmits the superposition
\begin{equation}
X_{t}=\sqrt{\alpha P}X_{1,t}+\sqrt{\bar{\alpha}P}X_{2,t},\label{eq:BC mode}
\end{equation}
for each symbol, where $\alpha\in[0,1]$ and $\bar{\alpha}=1-\alpha$
represent the fractions of powers allocated to the first and second
layers, respectively. In contrast, in the SL mode, the encoder transmits
only the second-layer codeword with full power $P$, and the transmitted
signal $X_{t}$ is written as
\begin{equation}
X_{t}=\sqrt{P}X_{2,t}.\label{eq:SL mode}
\end{equation}

In the first slot $t=1$, the encoder emits the signal $X_{1}$ in
the BC mode (\ref{eq:BC mode}) and the relay sends the compressed
version $W_{1}$ in (\ref{eq:Gaussian test channel}) of the received
signal $V_{1}$ to the decoder. At the completion of the slot, the
decoder first tries to decode the message $M_{1}$; if successful,
it cancels the codeword $X_{1,1}$ from the received signal and attempts
to decode message $M_{2}$. Decoding is based on the received signal
in slot $t$, which can be written as
\begin{equation}
\bar{Y}_{t}=\left[\begin{array}{c}
W_{t}\\
Y_{t}
\end{array}\right]=\left[\begin{array}{c}
\sqrt{a_{t}D_{t}}\\
\sqrt{S_{t}}
\end{array}\right]X_{t}+\left[\begin{array}{c}
\sqrt{a_{t}D_{t}}E_{t}+N_{t}\\
Z_{t}
\end{array}\right],
\end{equation}
 for $t=1,\ldots,T$.

The decoder informs the encoder and the relay about the number of
layers that were correctly decoded. If both messages are not correctly
decoded, in the next slot $t=2$, the encoder sends incremental redundancy
information for both layers using the BC mode (\ref{eq:BC mode}).
Note that incremental redundancy entails that, as mentioned, the codebooks
used at different blocks are independent (see, e.g., \cite{Caire}).
Instead, if only the first layer $M_{1}$ was decoded in the first
slot, the encoder transmits incremental redundancy information only
for the second layer by using the SL mode (\ref{eq:SL mode}). This
process lasts until either both messages $M_{1}$ and $M_{2}$ are
decoded successfully or the maximum number $T$ of transmissions is
reached. Therefore, if a message $M_{l}$ is not decoded until the
$T$th slot, outage is declared for layer $l$.

\section{Problem Definition\label{sec:Problem-Definition}}

The problem of interest is the maximization of the expected throughput
$\eta$ as measured by the average number of successfully received
bits per channel use. Using the renewal theorem (see, e.g., \cite{Zorzi}),
we can calculate the expected throughput $\eta$ as
\begin{equation}
\eta=\frac{\mathbb{E}\left[\mathtt{R}\right]}{\mathbb{E}\left[\mathtt{L}\right]},\label{eq:average throughput}
\end{equation}
where $\mathbb{E}\left[\mathtt{R}\right]$ is the average rate decoded
in a HARQ session, which consists of at most $T$ transmissions, and
$\mathbb{E}\left[\mathtt{L}\right]$ is the expected number of transmission
blocks for HARQ session. Expectations are taken with respect to the
fading coefficients $D_{t}$ and $S_{t}$. These quantities can be
computed as \cite{Steiner:HARQ}
\begin{align}
\mathbb{E}\left[\mathtt{R}\right] & =R_{1}\left(1-p_{\mathrm{out}}^{1}(T)\right)+R_{2}\left(1-p_{\mathrm{out}}^{2}(T)\right),\label{eq:average reward}\\
\mathrm{and}\,\,\mathbb{E}\left[\mathtt{L}\right] & =\sum_{t=1}^{T-1}tp_{\mathrm{dec}}^{2}(t)+T\left(p_{\mathrm{dec}}^{2}(T)+p_{\mathrm{out}}^{2}(T)\right),\label{eq:average delay}
\end{align}
where the probabilities $p_{\mathrm{out}}^{l}(k)$ and $p_{\mathrm{dec}}^{l}(k)$
are defined as
\begin{align}
p_{\mathrm{out}}^{l}(k) & =\Pr\left[M_{l}\,\mathrm{is\, not\, decoded\, until\, slot}\, k\right],\label{eq:probability outage}\\
\mathrm{and}\,\, p_{\mathrm{dec}}^{l}(k) & =\Pr\left[M_{l}\,\mathrm{is\, decoded\, in\, slot}\, k\right].\label{eq:probability successful decoding}
\end{align}
The probabilities $p_{\mathrm{out}}^{l}(k)$ and $p_{\mathrm{dec}}^{l}(k)$
depend on the parameters $R_{1}$, $R_{2}$ and $\alpha$ as will
be clarified in the next sections. The problem of maximizing the average
throughput $\eta$ is then formulated as
\begin{equation}
\underset{R_{1},R_{2}\geq0,\,\alpha\in[0,1]}{\mathrm{maximize}}\,\,\eta(R_{1},R_{2},\alpha),\label{eq:problem BC}
\end{equation}
where we have made explicit the dependence on $(R_{1},R_{2},\alpha)$.
As a benchmark, it is useful to consider the single-layer scheme obtained
as a special case of the proposed strategy with $R_{2}=0$ and $\alpha=1$.
Thus, the optimal throughput of a single-layer strategy is the solution
of the following problem:
\begin{equation}
\underset{R_{1}\geq0}{\mathrm{maximize}}\,\,\eta(R_{1},0,1).\label{eq:problem single-layer}
\end{equation}

\section{Constant Compression Gain\label{sec:fixed compression}}

In this section, we analyze the throughput of the proposed multi-layer
HARQ strategy when the relay uses a constant compression gain $a_{t}=a_{D_{t}}$
as in (\ref{eq:compression gain conservative}) for all $t=1,\ldots,T$
regardless of the feedback information reported from the decoder.
As explained in Sec. \ref{sec:System-Model}, with this choice, the
description $W_{t}$ can be recovered at the decoder for all realizations
of the fading channel $S_{t}$. However, this approach is not able
to opportunistically leverage a more advantageous fading state $S_{t}$.
A strategy that can exploit better fading state via adaptive compression
will be discussed in Sec. \ref{sec:Adaptive Compression}. We focus
on the LTSC model, so that $D_{t}=D$ and $S_{t}=S$ for all $t=1,\ldots,T$.
Moreover, we study both the case with \textit{local CSI} at the encoder,
i.e., when the encoder knows the local fading state $D=d$ and thus
can choose the tuple $(R_{1}(d),R_{2}(d),\alpha(d))$ as a function
of $d$, and the case with no local CSI at the encoder.

To express the objective throughput $\eta$ in (\ref{eq:average throughput}),
we have to compute the probabilities in (\ref{eq:probability outage})
and (\ref{eq:probability successful decoding}) as a function of parameters
$R_{1}$, $R_{2}$ and $\alpha$ which is done in the following lemmas.

\begin{lemma}\label{lem:p1 out LCSIT} The probability $p_{\mathrm{out}}^{1}(k)$
with compression gain $a_{D}$ is given as
\begin{align}
p_{\mathrm{out}}^{1}(k) & =\mathbb{E}_{D}\left[\theta(D)\right]\label{eq:p1 out local CSIT}
\end{align}
where the function $\theta(d)$ is defined as
\[
\theta(d)=\begin{cases}
F_{S}\left(\frac{\zeta_{1}(d,1)}{\bar{\zeta}_{1}(d,1)P}+\frac{a_{d}}{b_{d}}\right), & \mathrm{if}\,\,\bar{\zeta}_{1}(d,1)\leq0\\
1, & \mathrm{if}\,\,\bar{\zeta}_{1}(d,1)>0
\end{cases},
\]
with $b_{d}\triangleq1+a_{d}/d$ and the functions $\zeta_{i}(d,l)$
and $\bar{\zeta}_{i}(d,l)$ given as
\begin{align}
\zeta_{i}(d,l)= & 2^{2R_{i}(d)/l}-1,\\
\mathrm{and}\,\,\bar{\zeta}_{i}(d,l)= & 2^{2R_{i}(d)/l}\bar{\alpha}(d)-1,
\end{align}
for $i=1,2$ and $k=1,\ldots,T$.

\begin{proof} The proof is in Appendix \ref{appendix:p1 out LCSIT}.

\end{proof}

\end{lemma}

\begin{lemma}\label{lem:p2 out LCSIT} If $\bar{\alpha}P\ll1$, the
probability $p_{\mathrm{out}}^{2}(k)$ with compression gain $a_{D}$
is approximated as
\begin{align}
p_{\mathrm{out}}^{2}(k)\approx & p_{\mathrm{out}}^{1}(k)+\sum_{l=1}^{k}\mathbb{E}_{D}\left[\left[F_{S}(v_{k,l}^{\mathrm{UB}}(D))-F_{S}(v_{l}^{\mathrm{LB}}(D))\right]^{+}\right],\label{eq:p2 out local CSIT}
\end{align}
where $v_{l}^{\mathrm{LB}}(d)$ and $v_{k,l}^{\mathrm{UB}}(d)$ are
defined as
\begin{align}
v_{l}^{\mathrm{LB}}(d)= & \begin{cases}
-\left[-\frac{\zeta_{1}(d,l)}{\bar{\zeta}_{1}(d,l)P}-\frac{a_{d}}{b_{d}}\right]^{+}, & \mathrm{if}\,\bar{\zeta}_{1}(d,l)<0\\
\infty, & \mathrm{if}\,\bar{\zeta}_{1}(d,l)\geq0
\end{cases},\\
\mathrm{and}\,\, v_{k,l}^{\mathrm{UB}}(d)= & \min\mathcal{V}_{k,l}^{\mathrm{UB}},
\end{align}
with the set $\mathcal{V}_{k,l}^{\mathrm{UB}}$ given as
\begin{equation}
\mathcal{V}_{k,l}^{\mathrm{UB}}=\left\{ \hat{v}_{k,l}^{\mathrm{UB}}(d)\right\} \cup\begin{cases}
\textrm{Ø}, & \mathrm{if}\,\bar{\zeta}_{1}(d,l-1)\geq0\\
\left\{ \left[-\frac{\zeta_{1}(d,l-1)}{\bar{\zeta}_{1}(d,l-1)P}-\frac{a_{d}}{b_{d}}\right]^{+}\right\} , & \mathrm{if}\,\bar{\zeta}_{1}(d,l-1)<0
\end{cases}.
\end{equation}
We have defined the function $\hat{v}_{k,l}^{\mathrm{UB}}(d)$ as
\begin{equation}
\hat{v}_{k,l}^{\mathrm{UB}}(d)=\begin{cases}
\infty, & \mathrm{if}\,\, k=l\,\mathrm{and}\,\frac{2^{2R_{2}(d)}b_{d}^{k}}{c(d)^{l}}>1\\
0, & \mathrm{if}\,\, k=l\,\mathrm{and}\,\frac{2^{2R_{2}(d)}b_{d}^{k}}{c(d)^{l}}\leq1\\
\left[\frac{\left(\zeta_{2}(d,k-l)+1\right)b_{d}^{k/(k-l)-1}}{Pc(d)^{l/(k-l)}}-\frac{a_{d}}{b_{d}}-\frac{1}{P}\right]^{+}, & \mathrm{if}\,\, k>l
\end{cases},
\end{equation}
with the function $c(d)$ given as $c(d)=b_{d}+\bar{\alpha}(d)Pa_{d}$.

\begin{proof} See Appendix \ref{appendix:p2 out LCSIT}.

\end{proof}

\end{lemma}

\begin{lemma}\label{lem:p2 dec LCSIT} If $\bar{\alpha}P\ll1$, the
probability $p_{\mathrm{dec}}^{2}(k)$ with compression gain $a_{D}$
is approximated as
\begin{align}
p_{\mathrm{dec}}^{2}(k)\approx & \sum_{l=1}^{k-1}\mathbb{E}_{D}\left[\left[F_{S}(u_{k,l}^{\mathrm{UB}}(D))-F_{S}(u_{k,l}^{\mathrm{LB}}(D))\right]^{+}\right]+\mathbb{E}_{D}\left[\left[F_{S}(q_{k}^{\mathrm{UB}}(D))-F_{S}(q_{k}^{\mathrm{LB}}(D))\right]^{+}\right],\label{eq:p2 dec local CSIT}
\end{align}
where $u_{k,l}^{\mathrm{LB}}(d)$, $u_{k,l}^{\mathrm{UB}}(d)$, $q_{k}^{\mathrm{LB}}(d)$
and $q_{k}^{\mathrm{UB}}(d)$ are defined as
\begin{align*}
u_{k,l}^{\mathrm{LB}}(d)= & \max\mathcal{U}_{k,l}^{\mathrm{LB}},\\
u_{k,l}^{\mathrm{UB}}(d)= & \min\mathcal{U}_{k,l}^{\mathrm{UB}},\\
q_{k}^{\mathrm{LB}}(d)= & \max\mathcal{Q}_{k}^{\mathrm{LB}},\\
\mathrm{and}\,\, q_{k}^{\mathrm{UB}}(d)= & \begin{cases}
-\frac{\zeta_{1}(d,k-1)}{\bar{\zeta}_{1}(d,k-1)P}-\frac{a_{d}}{b_{d}}, & \mathrm{if}\,\bar{\zeta}_{1}(d,k-1)<0\\
\infty, & \mathrm{if}\,\bar{\zeta}_{1}(d,k-1)\geq0
\end{cases},
\end{align*}
with the sets $\mathcal{U}_{k,l}^{\mathrm{LB}}$, $\mathcal{U}_{k,l}^{\mathrm{UB}}$
and $\mathcal{Q}_{k}^{\mathrm{LB}}$ given as
\begin{align}
\mathcal{U}_{k,l}^{\mathrm{LB}}= & \left\{ \hat{u}_{k,l}^{\mathrm{LB}}(d)\right\} \cup\begin{cases}
\left\{ -\frac{\zeta_{1}(d,l)}{\bar{\zeta}_{1}(d,l)P}-\frac{a_{d}}{b_{d}}\right\} , & \mathrm{if}\,\bar{\zeta}_{1}(d,l)<0\\
\{\infty\}, & \mathrm{if}\,\bar{\zeta}_{1}(d,l)\geq0
\end{cases},\\
\mathcal{U}_{k,l}^{\mathrm{UB}}= & \left\{ \hat{u}_{k,l}^{\mathrm{UB}}(d)\right\} \cup\begin{cases}
\left\{ -\frac{\zeta_{1}(d,l-1)}{\bar{\zeta}_{1}(d,l-1)P}-\frac{a_{d}}{b_{d}}\right\} , & \mathrm{if}\,\bar{\zeta}_{1}(d,l-1)<0\\
\textrm{Ø}, & \mathrm{if}\,\bar{\zeta}_{1}(d,l-1)\geq0
\end{cases},\\
\mathrm{and}\,\,\mathcal{Q}_{k}^{\mathrm{LB}}= & \left\{ \left[\frac{\zeta_{2}(d,k)}{\bar{\alpha}(d)P}-\frac{a_{d}}{b_{d}}\right]^{+}\right\} \cup\begin{cases}
\left\{ -\frac{\zeta_{1}(d,k)}{\bar{\zeta}_{1}(d,k)P}-\frac{a_{d}}{b_{d}}\right\} , & \mathrm{if}\,\bar{\zeta}_{1}(d,k)<0\\
\{\infty\}, & \mathrm{if}\,\bar{\zeta}_{1}(d,k)\geq0
\end{cases}.\nonumber
\end{align}
We have defined the functions $\hat{u}_{k,l}^{\mathrm{LB}}(d)$ and
$\hat{u}_{k,l}^{\mathrm{UB}}(d)$ as
\begin{align}
\hat{u}_{k,l}^{\mathrm{LB}}(d)= & \left[\frac{\left(\zeta_{2}(d,k-1)+1\right)b_{d}^{k/(k-1)-1}}{Pc(d)^{l/(k-l)}}-\frac{a_{d}}{b_{d}}-\frac{1}{P}\right]^{+},\\
\hat{u}_{k,l}^{\mathrm{UB}}(d)= & \begin{cases}
\infty, & \mathrm{if}\, l=k-1\,\mathrm{and}\,\frac{2^{2R_{2}(d)}b_{d}^{k-1}}{c(d)^{l}}>1\\
0 & \mathrm{if}\, l=k-1\,\mathrm{and}\,\frac{2^{2R_{2}(d)}b_{d}^{k-1}}{c(d)^{l}}\leq1\\
\left[\frac{\left(\zeta_{2}(d,k-1-l)+1\right)b_{d}^{(k-1)/(k-1-l)-1}}{Pc(d)^{l/(k-1-l)}}-\frac{a_{d}}{b_{d}}-\frac{1}{P}\right]^{+}, & \mathrm{if}\, l<k-1
\end{cases}.
\end{align}

\begin{proof} The proof appears in Appendix \ref{appendix:p2 dec LCSIT}.

\end{proof}

\end{lemma}

With Lemmas \ref{lem:p1 out LCSIT}-\ref{lem:p2 dec LCSIT}, we can
express the throughput (\ref{eq:average throughput}) as a function
of the optimization variables $R_{1}$, $R_{2}$ and $\alpha$ via
numerical integration over the distribution $F_{D}(d)$. The optimization
problems (\ref{eq:problem BC}) and (\ref{eq:problem single-layer})
are not convex and need to be solved via global optimization tool
such as the branch-and-bound method \cite{Bertsekas}. Specifically,
with local CSI at the encoder, one needs to optimize over the parameters
$R_{1}(d)$, $R_{2}(d)$ and $\alpha(d)$, which corresponds to the
tuple $(R_{1},R_{2},\alpha)$ to be used when the relay fading state
is $D_{t}=d$. In practice, this optimization can be reformulated
by quantizing the fading distribution. Instead, without local CSI
at the encoder, the optimization is done over the single tuple $(R_{1},R_{2},\alpha)$
since the encoder is not able to adapt to the fading state $D$.

\section{Adaptive Compression Gain\label{sec:Adaptive Compression}}

In the previous section, we have assumed that the relay employs Gaussian
test channel (\ref{eq:Gaussian test channel}) with compression gain
$a_{t}=a_{D}$ for all $t=1,\ldots,T$ regardless of the feedback
information reported from the decoder under the LTSC (\ref{eq:LTSC model}).
We recall that this choice guarantees reliable decompression even
in the worst-case fading state $S_{t}=s_{\min}$. This section is
motivated by the attempt to leverage better fading states when they
occur. To this end, we assume that the feedback information that only
the message $M_{1}$ of the first layer was decoded in a slot $k$
implies that the fading coefficient $S_{t}$ of the side information
$Y_{t}$ is larger than some level $\hat{s}$, that is, $S\geq\hat{s}$.
This can be calculated as
\begin{equation}
\hat{s}=\max\left\{ \frac{\left[(2^{2R_{1}/k}-1)b_{D}+(2^{2R_{1}/k}\bar{\alpha}-1)Pa_{D}\right]^{+}}{b_{D}(1-2^{2R_{1}/k}\bar{\alpha})P},\, s_{\min}\right\} \label{eq:adaptive comp s dash}
\end{equation}
by imposing the condition that the accumulated mutual information
is sufficient to support rate $R_{1}$ (see Appendix \ref{Appendix I}
for more discussion). Upon reception of a positive acknowledgement
for layer 1 and a negative acknowledgement for layer 2, we then propose
that, from the next slot $t>k$, the relay performs compression assuming
the better side information $S_{t}=\hat{s}\geq s_{\min}$. The corresponding
compression gain is given as
\begin{equation}
a_{t}=\frac{\beta(1+\hat{s}P)}{1/D+(1+\hat{s}/D)P}\triangleq\hat{a}_{D}.
\end{equation}

With adaptive compression, the expected throughput $\eta$ in (\ref{eq:average throughput})
can be computed using the lemmas presented in the previous section
with the changes discussed in the following lemmas.

\begin{lemma}\label{lem:p2 out adaptive compression}

If $\bar{\alpha}P\ll1$, the probability $p_{\mathrm{out}}^{2}(k)$
with adaptive compression is approximated as (\ref{eq:p2 out local CSIT})
in Lemma \ref{lem:p2 out LCSIT} with $\hat{v}_{k,l}^{\mathrm{UB}}(d)$
modified as
\begin{equation}
\hat{v}_{k,l}^{\mathrm{UB}}(d)=\begin{cases}
\infty, & \mathrm{if}\,\, k=l\,\mathrm{and}\,\frac{2^{2R_{2}(d)}b_{d}^{l}}{c(d)^{l}}>1\\
0, & \mathrm{if}\,\, k=l\,\mathrm{and}\,\frac{2^{2R_{2}(d)}b_{d}^{l}}{c(d)^{l}}\leq1\\
\left[\frac{\left(\zeta_{2}(d,k-l)+1\right)b_{d}^{l/(k-l)}}{Pc(d)^{l/(k-l)}}-\frac{\hat{a}_{d}}{\hat{b}_{d}}-\frac{1}{P}\right]^{+}, & \mathrm{if}\,\, k>l
\end{cases},
\end{equation}
where $\hat{b}_{d}\triangleq1+\hat{a}_{d}/d$.

\begin{proof}See Appendix \ref{appendix:p2 out adaptive compression}.

\end{proof}

\end{lemma}

\begin{lemma}\label{lem:p2 dec adaptive compression}

If $\bar{\alpha}P\ll1$, the probability $p_{\mathrm{dec}}^{2}(k)$
with adaptive compression is approximated as (\ref{eq:p2 dec local CSIT})
in Lemma \ref{lem:p2 dec LCSIT} with $\hat{u}_{k,l}^{\mathrm{LB}}(d)$
and $\hat{u}_{k,l}^{\mathrm{UB}}(d)$ modified as
\begin{align}
\hat{u}_{k,l}^{\mathrm{LB}}(d)= & \left[\frac{\left(\zeta_{2}(d,k-l)+1\right)b_{d}^{l/(k-l)}}{Pc(d)^{l/(k-l)}}-\frac{\hat{a}_{d}}{\hat{b}_{d}}-\frac{1}{P}\right]^{+},\\
\hat{u}_{k,l}^{\mathrm{UB}}(d)= & \begin{cases}
\infty, & \mathrm{if}\,\, l=k-1\,\mathrm{and}\,\frac{2^{2R_{2}(d)}b_{d}^{l}}{c(d)^{l}}>1\\
0, & \mathrm{if}\,\, l=k-1\,\mathrm{and}\,\frac{2^{2R_{2}(d)}b_{d}^{l}}{c(d)^{l}}\leq1\\
\left[\frac{\left(\zeta_{2}(d,k-1-l)+1\right)b_{d}^{l/(k-1-l)}}{Pc(d)^{l/(k-1-l)}}-\frac{\hat{a}_{d}}{\hat{b}_{d}}-\frac{1}{P}\right]^{+}, & \mathrm{if}\,\, l<k-1
\end{cases}.
\end{align}

\begin{proof}See Appendix \ref{appendix:p2 dec adaptive compression}.

\end{proof}

\end{lemma}

With the results in Lemmas \ref{lem:p1 out LCSIT}, \ref{lem:p2 out adaptive compression}
and \ref{lem:p2 dec adaptive compression}, we can express the average
throughput with the adaptive compression described in this section
as a function of the design parameters $R_{1}$, $R_{2}$ and $\alpha$.

\section{Short-Term Static Channels\label{sec:STSC model}}

In this section, we discuss the STSC model in which the channel coefficients
$D_{t}$ and $S_{t}$ change independently from block to block. For
simplicity, as in \cite{Steiner:HARQ}, we focus on the case $T=2$,
i.e., there can be at most one retransmission. It is observed that,
even with $T=2$, we have to consider four random variables $D_{1}$,
$D_{2}$, $S_{1}$ and $S_{2}$, which complicate the analysis as
compared to the LTSC model. Moreover, given the independence of the
channel fading gains from block to block, adaptive compression is
not applicable under the STSC model. Therefore, we set the compression
gains as $a_{t}=a_{D_{t}}$ in (\ref{eq:compression gain conservative})
for $t=1,2$. The quantities in (\ref{eq:average reward}) and (\ref{eq:average delay})
reduce to
\begin{align}
\mathbb{E}\left[\mathtt{R}\right] & =R_{1}\left(1-p_{\mathrm{out}}^{1}(2)\right)+R_{2}\left(1-p_{\mathrm{out}}^{2}(2)\right),\\
\mathrm{and}\,\,\mathbb{E}\left[\mathtt{L}\right] & =p_{\mathrm{dec}}^{2}(1)+2\left(1-p_{\mathrm{dec}}^{2}(1)\right).
\end{align}
Thus, it is enough to compute three probabilities $p_{\mathrm{out}}^{1}(2)$,
$p_{\mathrm{out}}^{2}(2)$ and $p_{\mathrm{dec}}^{2}(1)$, which are
derived in the following lemmas.

\begin{lemma}\label{lem:p1 out 2 STSC} The probability $p_{\mathrm{out}}^{1}(2)$
in the STSC model with $T=2$ is given as
\begin{equation}
p_{\mathrm{out}}^{1}(2)=\mathbb{E}_{D_{1},D_{2},S_{1}}\left[g(D_{1},D_{2},S_{1})\right],\label{eq:p1 out 2 STSC}
\end{equation}
where we have defined the function $g(d_{1},d_{2},s_{1})$ as
\begin{align}
g(d_{1},d_{2},s_{1}) & =\begin{cases}
F_{S_{2}}\left(\frac{\left[(2^{2h(d_{1},s_{1})}-1)b_{d_{2}}+(2^{2h(d_{1},s_{1})}\bar{\alpha}-1)a_{d_{2}}P\right]^{+}}{(2^{2h(d_{1},s_{1})}\bar{\alpha}-1)b_{d_{2}}P}\right), & \mathrm{if}\,\,2^{2h(d_{1},s_{1})}\bar{\alpha}<1\\
0, & \mathrm{if}\,\, h(d_{1},s_{1})=0\,\mathrm{and}\,\bar{\alpha}=1\\
1, & \mathrm{otherwise}
\end{cases},
\end{align}
with the function $h(d_{1},s_{1})$ given as
\begin{align}
h(d_{1},s_{1})= & R_{1}-f_{\mathrm{I}}\left(\alpha P,\bar{\alpha}P,a_{d_{1}},s_{1},d_{1}\right).
\end{align}
The function $f_{\mathrm{I}}(P,\bar{P},a,s,d)$ is defined as
\begin{equation}
f_{\mathrm{I}}(P,\bar{P},a,s,d)=\frac{1}{2}\log\left(1+P\frac{s+a(1+s/d)}{1+a/d+\bar{P}\left(s+a(1+s/d)\right)}\right).\label{eq:function mutual information}
\end{equation}

\begin{proof} See Appendix \ref{Appendix:proof for STSC}.

\end{proof}

\end{lemma}

\begin{lemma}\label{lem:p2 out 2 STSC} The probability $p_{\mathrm{out}}^{2}(2)$
in the STSC model with $T=2$ is given as
\begin{align}
p_{\mathrm{out}}^{2}(2)= & \mathbb{E}_{D_{1},D_{2},S_{1}}\left[\varphi(D_{1},D_{2},S_{1})\right]+\mathbb{E}_{D_{1},D_{2},S_{1}}\left[\gamma(D_{1},D_{2},S_{1})\right]+p_{\mathrm{out}}^{1}(2),\label{eq:p2 out 2 STSC}
\end{align}
where the functions $\varphi(d_{1},d_{2},s_{1})$ and $\gamma(d_{1},d_{2},s_{1})$
are defined as
\begin{align}
\varphi(d_{1},d_{2},s_{1}) & =\begin{cases}
F_{S_{2}}\left(\left[\frac{2^{2\left(R_{2}-f_{\mathrm{I}}\left(\bar{\alpha}P,0,a_{d_{1}},s_{1},d_{1}\right)\right)}-1}{P}-\frac{a_{d_{2}}}{b_{d_{2}}}\right]^{+}\right), & \mathrm{if}\,\, f_{\mathrm{I}}\left(\alpha P,\bar{\alpha}P,a_{d_{1}},s_{1},d_{1}\right)\geq R_{1}\\
0, & \mathrm{if}\,\, f_{\mathrm{I}}\left(\alpha P,\bar{\alpha}P,a_{d_{1}},s_{1},d_{1}\right)<R_{1}
\end{cases},\\
\mathrm{and}\,\,\gamma(d_{1},d_{2},s_{1}) & =\begin{cases}
\left[F_{S_{2}}\left(\mu_{\mathrm{UB}}(s_{1})\right)-F_{S_{2}}\left(\mu_{\mathrm{LB}}(s_{1})\right)\right]^{+}, & \mathrm{if}\,\, f_{\mathrm{I}}\left(\alpha P,\bar{\alpha}P,a_{d_{1}},s_{1},d_{1}\right)<R_{1}\\
0, & \mathrm{if}\,\, f_{\mathrm{I}}\left(\alpha P,\bar{\alpha}P,a_{d_{1}},s_{1},d_{1}\right)\geq R_{1}
\end{cases},
\end{align}
with $\mu_{\mathrm{UB}}(s_{1})$ and $\mu_{\mathrm{LB}}(s_{1})$ given
as
\begin{align}
\mu_{\mathrm{UB}}(s_{1}) & =\begin{cases}
\infty, & \mathrm{if}\,\,\bar{\alpha}=0\,\mathrm{and}\,2^{2\left(R_{2}-f_{\mathrm{I}}\left(\bar{\alpha}P,0,a_{d_{1}},s_{1},d_{1}\right)\right)}>1\\
0, & \mathrm{if}\,\,\bar{\alpha}=0\,\mathrm{and}\,2^{2\left(R_{2}-f_{\mathrm{I}}\left(\bar{\alpha}P,0,a_{d_{1}},s_{1},d_{1}\right)\right)}\leq1\\
\frac{\left[\left(2^{2\left(R_{2}-f_{\mathrm{I}}\left(\bar{\alpha}P,0,a_{d_{1}},s_{1},d_{1}\right)\right)}-1\right)b_{d_{2}}-\bar{\alpha}Pa_{d_{2}}\right]^{+}}{b_{d_{2}}\bar{\alpha}P}, & \mathrm{if}\,\,\bar{\alpha}>0
\end{cases},\\
\mathrm{and}\,\,\mu_{\mathrm{LB}}(S_{1}) & =\begin{cases}
\frac{\left[\left(2^{2\left(R_{1}-f_{\mathrm{I}}\left(\alpha P,\bar{\alpha}P,a_{d_{1}},s_{1},d_{1}\right)\right)}-1\right)b_{d_{2}}-\alpha Pa_{d_{2}}\right]^{+}}{\left(1-2^{2\left(R_{1}-f_{\mathrm{I}}\left(\alpha P,\bar{\alpha}P,a_{d_{1}},s_{1},d_{1}\right)\right)}\bar{\alpha}\right)b_{d_{2}}P}, & \mathrm{if}\,\,2^{2\left(R_{1}-f_{\mathrm{I}}\left(\alpha P,\bar{\alpha}P,a_{d_{1}},s_{1},d_{1}\right)\right)}\bar{\alpha}<1\\
0, & \mathrm{if}\,\,2^{2\left(R_{1}-f_{\mathrm{I}}\left(\alpha P,\bar{\alpha}P,a_{d_{1}},s_{1},d_{1}\right)\right)}\bar{\alpha}=1\\
 & \,\,\,\,\mathrm{and}\,\frac{\alpha b_{d_{2}}}{\bar{\alpha}}\leq\alpha Pa_{d_{2}}\\
\infty & \mathrm{otherwise}
\end{cases}.
\end{align}

\begin{proof} See Appendix \ref{Appendix:proof for STSC}.

\end{proof}

\end{lemma}

\begin{lemma}\label{lem:p2 dec 1 STSC} The probability $p_{\mathrm{dec}}^{2}(1)$
in the STSC model with $T=2$ is given as
\begin{equation}
p_{\mathrm{dec}}^{2}(1)=\mathbb{E}_{D_{1}}\left[\max\left\{ \lambda_{1}(D_{1}),\lambda_{2}(D_{1})\right\} \right],\label{eq:p2 dec 1 STSC}
\end{equation}
with the functions $\lambda_{1}(d_{1})$ and $\lambda_{2}(d_{1})$
given as
\begin{align*}
\lambda_{1}(D_{1}) & =\begin{cases}
\frac{\left[(2^{2R_{1}}-1)b_{d_{1}}+(2^{2R_{1}}\bar{\alpha}-1)Pa_{d_{1}}\right]^{+}}{(1-2^{2R_{1}}\bar{\alpha})b_{d_{1}}P}, & \mathrm{if}\,\,2^{2R_{1}}\bar{\alpha}<1\\
\infty, & \mathrm{if}\,\,2^{2R_{1}}\bar{\alpha}\geq1
\end{cases},\\
\mathrm{and}\,\,\lambda_{2}(D_{1}) & =\begin{cases}
\frac{\left[(2^{2R_{2}}-1)b_{d_{1}}-\bar{\alpha}Pa_{d_{1}}\right]^{+}}{b_{d_{1}}\bar{\alpha}P}, & \mathrm{if}\,\,\bar{\alpha}>0\\
\infty, & \mathrm{if}\,\,\bar{\alpha}=0\,\mathrm{and}\, R_{2}>0\\
0, & \mathrm{if}\,\,\bar{\alpha}=0\,\mathrm{and}\, R_{2}=0
\end{cases}.
\end{align*}

\begin{proof} See Appendix \ref{Appendix:proof for STSC}.

\end{proof}

\end{lemma}

\section{Numerical Results\label{sec:Numerical-Results}}

\begin{figure}
\centering\includegraphics[width=12cm,height=9cm]{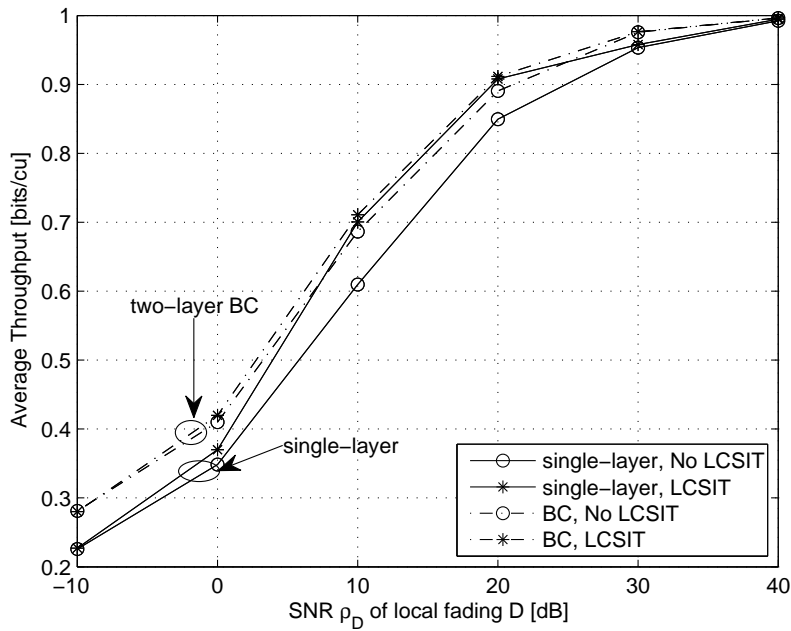}

\caption{\label{fig:graph SNR_D}Average throughput $\eta$ versus the SNR
$\rho_{D}$ with $T=2$, $C_{\max}=1$, $P=0\,\mathrm{dB}$, $\rho_{S}=0\,\mathrm{dB}$
and $K=0$.}
\end{figure}

In this section, we present numerical results to gain insights into
the advantage of the proposed multi-layer HARQ strategies. In the
figures, the cases with and without local CSI at the encoder are denoted
by ``LCSIT'' and ``No LCSIT'', respectively. We assume Rayleigh
fading for the side information $Y_{t}$ and Rician fading for the
signal $V_{t}$ received by the relay with Rician factor $K$ (i.e.,
$K$ is the ratio of the power of line-of-sight (LOS) component to
that of non-LOS component). The rationale behind these distributions
comes from the application to the cloud radio access scenario (see
Sec. \ref{sec:Introduction}), in which $Y_{t}$ is the signal received
by out-of-cell BSs, which typically lack the direct LOS component,
unlike the signal $V_{t}$ received by the in-cell BS. The signal-to-noise
ratios (SNRs) of the source-to-relay and the source-to-destination
links are defined as $\mathbb{E}[|D_{t}|^{2}]=\rho_{D}$ and $\mathbb{E}[|S_{t}|^{2}]=\rho_{S}$,
respectively.

We first examine in Fig. \ref{fig:graph SNR_D} how the SNR $\rho_{D}$
of the relay fading channel $D$ affects the average throughput $\eta$
by plotting $\eta$ versus the SNR $\rho_{D}$ under the LTSC model
with $T=2$, $C_{\max}=1$, $P=0\,\mathrm{dB}$, $\rho_{S}=0\,\mathrm{dB}$
and $K=0$. With local CSI at the encoder, the proposed BC scheme
shows performance gain over the conventional single-layer approach
only in the range of low SNR $\rho_{D}$. This is because in this
case, BC is only used to combat the uncertainty of the fading gain
$S_{t}$, whose relevance becomes less pronounced as $\rho_{D}$ increases.
However, with no local CSI at the encoder, the gain of the BC remains
substantial for all SNRs $\rho_{D}$, since in this case, the CSI
uncertainty at the encoder includes both $D_{t}$ and $S_{t}$.

\begin{figure}
\centering\includegraphics[width=12cm,height=9cm]{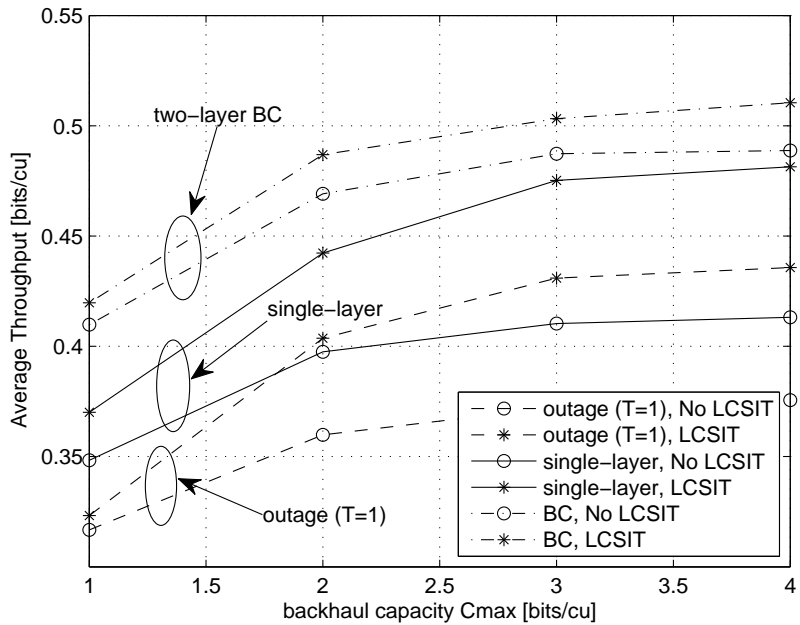}

\caption{\label{fig:graph Cmax}Average throughput $\eta$ versus the backhaul
capacity $C_{\max}$ with $T=2$, $P=0\,\mathrm{dB}$, $\rho_{D}=\rho_{S}=0\,\mathrm{dB}$
and $K=0$.}
\end{figure}

In Fig. \ref{fig:graph Cmax}, we plot the throughput performance
versus the backhaul capacity $C_{\max}$ for the LTSC model with $T=2$,
$P=0\,\mathrm{dB}$, $\rho_{D}=\rho_{S}=0\,\mathrm{dB}$ and $K=0$.
It is observed that the impact of the local CSI at the encoder becomes
more significant for larger backhaul capacities $C_{\max}$, since
the performance is more affected by the encoder-to-relay link if the
backhaul capacity $C_{\max}$ is large enough. Moreover, the flexibility
afforded by BC makes the effect of LCSIT less relevant than for conventional
single-layer transmission.

\begin{figure}
\centering\includegraphics[width=12cm,height=9cm]{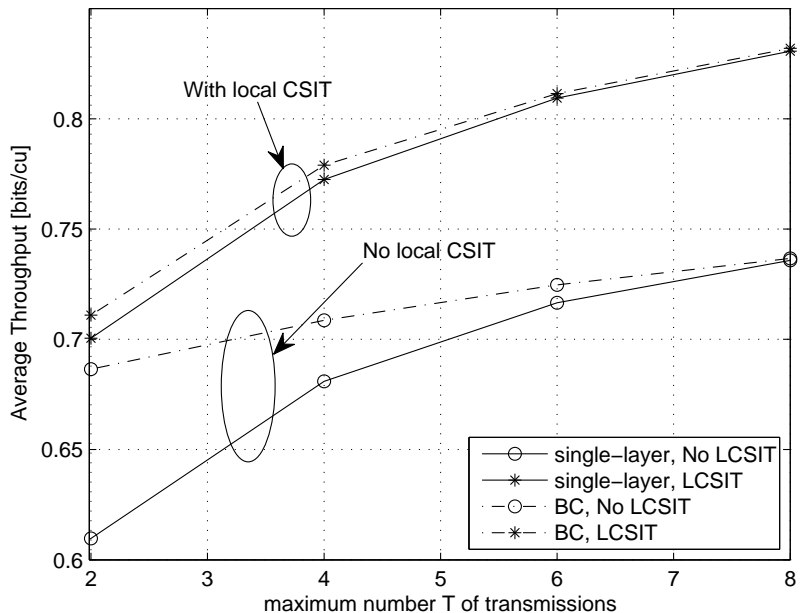}

\caption{\label{fig:graph T}Average throughput $\eta$ versus the maximum
number $T$ of transmissions with $C_{\max}=1$, $P=0\,\mathrm{dB}$,
$\rho_{D}=10\,\mathrm{dB}$, $\rho_{S}=0\,\mathrm{dB}$ and $K=0$.}
\end{figure}

In Fig. \ref{fig:graph T}, we observe the effect of the maximum number
$T$ of transmissions for the LTSC model with $C_{\max}=1$, $P=0\,\mathrm{dB}$,
$\rho_{D}=10\,\mathrm{dB}$, $\rho_{S}=0\,\mathrm{dB}$ and $K=0$.
For both cases with local CSI at the encoder or not, the advantage
of the BC scheme diminishes as $T$ increases. This implies that the
HARQ strategy is able to compensate for a large fraction of performance
degradation of the single-layer scheme when enough number of transmissions
are allowed. This trend is more apparent in the case with no local
CSI at the encoder, due to the layer gains of BC.

\begin{figure}
\centering\includegraphics[width=12cm,height=9cm]{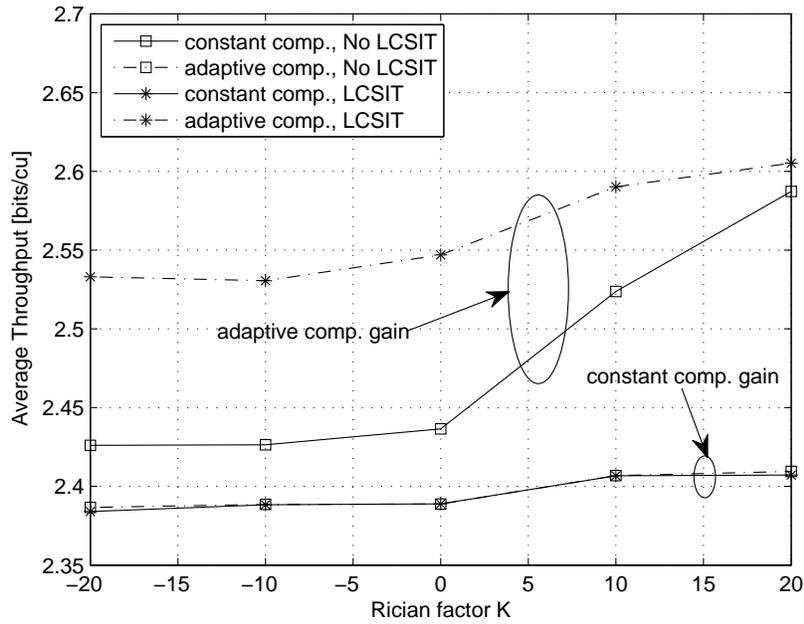}

\caption{\label{fig:graph adaptive comp}Average throughput $\eta$ versus
the factor $K$ of Rician fading with $C_{\max}=2$, $T=2$, $P=0\,\mathrm{dB}$
and $\rho_{D}=\rho_{S}=20\,\mathrm{dB}$.}
\end{figure}

In Fig. \ref{fig:graph adaptive comp}, we investigate the advantage
of the adaptive compression scheme proposed in Sec. \ref{sec:Adaptive Compression}
by plotting the throughput performance versus the Rician factor $K$
for the LTSC with $C_{\max}=2$, $T=2$, $P=0\,\mathrm{dB}$ and $\rho_{D}=\rho_{S}=20\,\mathrm{dB}$.
We recall that the adaptive compression was proposed to opportunistically
leverage better fading states. In accordance with this motivation,
the adaptive compression is observed to be advantageous as the factor
$K$ grows due to increased frequency of good fading states that can
be exploited via the proposed strategy.

\begin{figure}
\centering\includegraphics[width=12cm,height=9cm]{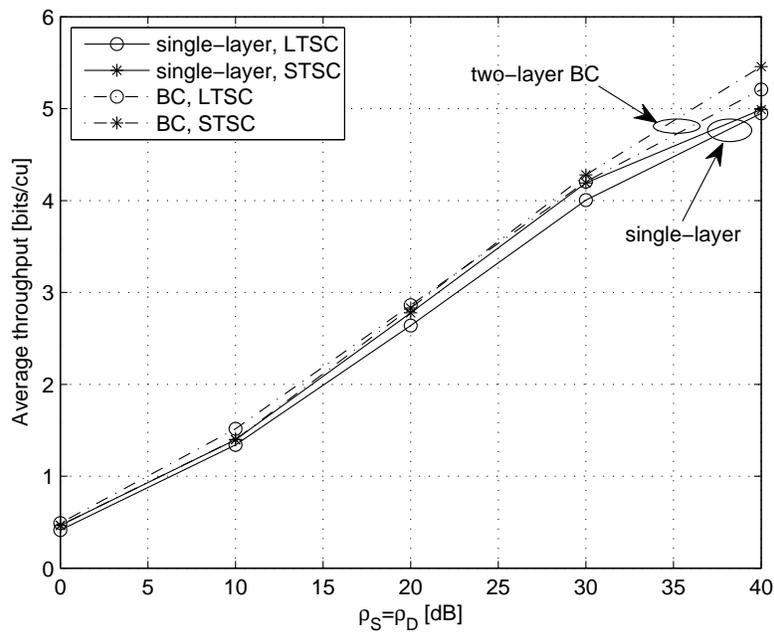}

\caption{\label{fig:graph STSC}Average throughput $\eta$ versus the SNRs
$\rho_{D}=\rho_{S}$ with no local CSI at the encoder and $T=2$,
$P=0\,\mathrm{dB}$, $C_{\max}=5$ and $K=0$.}
\end{figure}

Finally, in Fig. \ref{fig:graph STSC}, we compare the average throughput
performance under the LTSC and STSC models with no local CSI at the
encoder and $T=2$, $P=0\,\mathrm{dB}$, $C_{\max}=5$ and $K=0$.
With single-layer transmission, it is seen that the STSC model leads
to better performance than LTSC due to the diversity gain. However,
with BC, an additional factor determines the performance comparison,
namely the possibility for ``opportunistic retransmission'' under
LTSC. Specifically, under the LTSC model, when the encoder is reported
an ACK for the first-layer message $M_{1}$, it can transmit the second
layer $M_{2}$ in the next slot in order to leverage the good fading
state. In contrast, under the STSC model, this is not possible since
the fading coefficients $D_{t}$ and $S_{t}$ vary independently from
block to block. From the figure, it is observed that this factor is
dominant in the low-to-moderate SNR range, where the performance of
BC transmission under the LTSC model is better than under the STSC
model.

\section{Conclusions}

Motivated by the uplink of cloud radio access networks, we have studied
robust transmission and compression schemes for the fading relay channel
with an out-of-band relay. Specifically, we have adopted a multi-layer
BC transmission strategy coupled with HARQ, thus allowing for variable-length
transmission and variable-rate decoding, under two different channel
models, LTSC and STSC. Moreover, we have proposed an adaptive compression
strategy at the relay that is able to leverage better fading state
based on the HARQ feedback received from the destination. We have
demonstrated the performance gain of the proposed schemes over conventional
single-layer approaches via extensive simulations.

\appendices

\section{Proof of Lemmas \ref{lem:p1 out LCSIT}-\ref{lem:p2 dec LCSIT}}\label{Appendix I}

In this Appendix, we derive the probabilities presented in Lemmas
\ref{lem:p1 out LCSIT}-\ref{lem:p2 dec LCSIT}. Since we assumed the
LTSC model (\ref{eq:LTSC model}) in Sec. \ref{sec:fixed compression},
we have $D_{t}=D$ and $S_{t}=S$ for all $t=1,\ldots,T$. We first
calculate the probabilities conditioned on $D=d$ and the results
in Lemmas \ref{lem:p1 out LCSIT}-\ref{lem:p2 dec LCSIT} are then
obtained by taking expectation over $D$.

\subsection{Proof of Lemma \ref{lem:p1 out LCSIT}}\label{appendix:p1 out LCSIT}

In this subsection, we compute the probability $p_{\mathrm{out}}^{1}(k)$
that the message $M_{1}$ is not decoded until slot $k$. Since we
have assumed the IR-based HARQ approach, the probability $p_{\mathrm{out}}^{1}(k)$
can be calculated as the probability that the mutual information accumulated
along the first $k$ slots is smaller than $R_{1}$ \cite{Caire}:
\begin{align}
p_{\mathrm{out}}^{1}(k) & =\mathbb{E}_{D}\left[\Pr\left[k\cdot I(X_{1,t};\bar{Y}_{t}|\mathrm{BC\, mode})<R_{1}(D)|\, D\right]\right],\label{eq:Appendix I eq 1}
\end{align}
where the mutual information $I(X_{1,t};\bar{Y}_{t}|\mathrm{BC\, mode})$
is given as
\begin{equation}
I(X_{1,t};\bar{Y}_{t}|\mathrm{BC\, mode})=f_{\mathrm{I}}(\alpha(D)P,\bar{\alpha}(D)P,a_{D},S,D),\label{eq:mutual information X1 BC}
\end{equation}
with the function $f_{\mathrm{I}}(P,\bar{P},a,s,d)$ defined in (\ref{eq:function mutual information}).
If we express the probability (\ref{eq:Appendix I eq 1}) using the
CDF of $S$, we arrive at the expression (\ref{eq:p1 out local CSIT}).

\subsection{Proof of Lemma \ref{lem:p2 out LCSIT}}\label{appendix:p2 out LCSIT}

This subsection computes the probability $p_{\mathrm{out}}^{2}(k)$
that the message $M_{2}$ is not decoded until slot $k$. Using the
total probability theorem, we can write the probability $p_{\mathrm{out}}^{2}(k)$
as
\begin{align}
p_{\mathrm{out}}^{2}(k)= & \sum_{l=1}^{k}\mathbb{E}_{D}\left[\Pr\left[\begin{array}{c}
(l-1)\cdot I(X_{1,t};\bar{Y}_{t}|\mathrm{BC\, mode})<R_{1}(D)\leq l\cdot I(X_{1,t};\bar{Y}_{t}|\mathrm{BC\, mode}),\\
l\cdot I(X_{2,t};\bar{Y}_{t}|X_{1,t},\mathrm{BC\, mode})+(k-l)\cdot I(X_{2,t};\bar{Y}_{t}|X_{1,t},\mathrm{SL\, mode})<R_{2}
\end{array}\Bigg|\, D\right]\right]\nonumber \\
 & +p_{\mathrm{out}}^{1}(k),\label{eq:Appendix I eq 4}
\end{align}
where the probability $p_{\mathrm{out}}^{1}(k)$ was derived in the
previous subsection and the mutual information quantities related
to the second layer are given as
\begin{align}
I(X_{2,t};\bar{Y}_{t}|X_{1,t},\mathrm{BC\, mode})= & f_{\mathrm{I}}(\bar{\alpha}(D)P,\alpha P,a_{D},S,D),\label{eq:mutual information X2 BC}\\
I(X_{2,t};\bar{Y}_{t}|X_{1,t},\mathrm{SL\, mode})= & f_{\mathrm{I}}(P,0,a_{D},S,D).\label{eq:mutual information X2 SL}
\end{align}
The term inside the summation in (\ref{eq:Appendix I eq 4}) is then
derived as
\begin{align}
 & \Pr\left[\begin{array}{c}
(l-1)\cdot I(X_{1,t};\bar{Y}_{t}|\mathrm{BC\, mode})<R_{1}(D)\leq l\cdot I(X_{1,t};\bar{Y}_{t}|\mathrm{BC\, mode}),\\
l\cdot I(X_{2,t};\bar{Y}_{t}|X_{1,t},\mathrm{BC\, mode})+(k-l)\cdot I(X_{2,t};\bar{Y}_{t}|X_{1,t},\mathrm{SL\, mode})<R_{2}(D)
\end{array}\Bigg|\, D\right]\nonumber \\
= & \Pr\left[\begin{array}{c}
\left(1-2^{2R_{1}(D)/(l-1)}\bar{\alpha}(D)\right)b_{D}PS<(2^{2R_{1}(D)/(l-1)}-1)b_{D}+(2^{2R_{1}(D)/(l-1)}\bar{\alpha}(D)-1)Pa_{D},\\
(2^{2R_{1}(D)/l}-1)b_{D}+(2^{2R_{1}(D)/l}\bar{\alpha}(D)-1)Pa_{D}\leq(1-2^{2R_{1}(D)/l}\bar{\alpha}(D))b_{D}PS,\\
\left(b_{D}\left(1+\bar{\alpha}(D)PS\right)+\bar{\alpha}(D)Pa_{D}\right)^{l}\left(b_{D}(1+PS)+Pa_{D}\right)^{k-l}<2^{2R_{2}(D)}b_{D}^{k}
\end{array}\Bigg|D\right],\label{eq:Appendix I eq 2}
\end{align}
where the last condition makes it difficult to express the probability
in terms of the CDF of $S$. Following \cite{Steiner:HARQ}, we assume
the low SNR condition $\bar{\alpha}(D)P\ll1$ so that we have $1+\bar{\alpha}(D)PS\approx1$
in the last condition of the probability (\ref{eq:Appendix I eq 2}).
Then, the probability (\ref{eq:Appendix I eq 2}) is approximated
as
\begin{align}
 & \Pr\left[\begin{array}{c}
(l-1)\cdot I(X_{1,t};\bar{Y}_{t}|\mathrm{BC\, mode})<R_{1}(D)\leq l\cdot I(X_{1,t};\bar{Y}_{t}|\mathrm{BC\, mode}),\\
l\cdot I(X_{2,t};\bar{Y}_{t}|X_{1,t},\mathrm{BC\, mode})+(k-l)\cdot I(X_{2,t};\bar{Y}_{t}|X_{1,t},\mathrm{SL\, mode})<R_{2}(D)
\end{array}\Bigg|\, D\right]\nonumber \\
\approx & \left[F_{S}(v_{k,l}^{\mathrm{UB}}(D))-F_{S}(v_{k,l}^{\mathrm{LB}}(D))\right]^{+},\label{eq:Appendix I eq 3}
\end{align}
where $v_{k,l}^{\mathrm{UB}}(D)$ and $v_{k,l}^{\mathrm{LB}}(D)$
are defined in Lemma \ref{lem:p2 out LCSIT}. If we substitute (\ref{eq:Appendix I eq 3})
into (\ref{eq:Appendix I eq 4}), the result in Lemma \ref{lem:p2 out LCSIT}
is obtained.

\subsection{Proof of Lemma \ref{lem:p2 dec LCSIT}}\label{appendix:p2 dec LCSIT}

In this subsection, we compute the probability $p_{\mathrm{dec}}^{2}(k)$
that the message $M_{2}$ is successfully decoded in slot $k$. Following
similar arguments as above, we can write $p_{\mathrm{dec}}^{2}(k)$
as
\begin{align}
 & p_{\mathrm{dec}}^{2}(k)\nonumber \\
= & \sum_{l=1}^{k-1}\mathbb{E}_{D}\left[\Pr\left[\begin{array}{c}
(l-1)\cdot I(X_{1,t};\bar{Y}_{t}|\mathrm{BC\, mode})<R_{1}(D)\leq l\cdot I(X_{1,t};\bar{Y}_{t}|\mathrm{BC\, mode}),\\
l\cdot I(X_{2,t};\bar{Y}_{t}|X_{1,t},\mathrm{BC\, mode})+(k-1-l)\cdot I(X_{2,t};\bar{Y}_{t}|X_{1,t},\mathrm{SL\, mode})<R_{2}(D),\\
R_{2}(D)\leq l\cdot I(X_{2,t};\bar{Y}_{t}|X_{1,t},\mathrm{BC\, mode})+(k-l)\cdot I(X_{2,t};\bar{Y}_{t}|X_{1,t},\mathrm{SL\, mode})
\end{array}\Bigg|\, D\right]\right]\\
 & +\mathbb{E}_{D}\left[\Pr\left[\begin{array}{c}
(k-1)\cdot I(X_{1,t};\bar{Y}_{t}|\mathrm{BC\, mode})<R_{1}(D)\leq k\cdot I(X_{1,t};\bar{Y}_{t}|\mathrm{BC\, mode}),\\
R_{2}(D)\leq k\cdot I(X_{1,t};\bar{Y}_{t}|\mathrm{BC\, mode})
\end{array}\Bigg|\, D\right]\right].\label{eq:Appendix I eq 5}
\end{align}
Moreover, under the low SNR condition $\bar{\alpha}P\ll1$, the term
inside the summation is approximated as
\begin{align}
 & \Pr\left[\begin{array}{c}
(l-1)\cdot I(X_{1,t};\bar{Y}_{t}|\mathrm{BC\, mode})<R_{1}(D)\leq l\cdot I(X_{1,t};\bar{Y}_{t}|\mathrm{BC\, mode}),\\
l\cdot I(X_{2,t};\bar{Y}_{t}|X_{1,t},\mathrm{BC\, mode})+(k-1-l)\cdot I(X_{2,t};\bar{Y}_{t}|X_{1,t},\mathrm{SL\, mode})<R_{2}(D),\\
R_{2}(D)\leq l\cdot I(X_{2,t};\bar{Y}_{t}|X_{1,t},\mathrm{BC\, mode})+(k-l)\cdot I(X_{2,t};\bar{Y}_{t}|X_{1,t},\mathrm{SL\, mode})
\end{array}\Bigg|\, D\right]\nonumber \\
\approx & \left[F_{S}(u_{k,l}^{\mathrm{UB}}(D))-F_{S}(u_{k,l}^{\mathrm{LB}}(D))\right]^{+},\label{eq:Appendix I eq 6}
\end{align}
where we have defined $u_{k,l}^{\mathrm{UB}}(D)$ and $u_{k,l}^{\mathrm{LB}}(D)$
in Lemma \ref{lem:p2 dec LCSIT}. Moreover, we can derive the last
term in (\ref{eq:Appendix I eq 5}) as
\begin{align}
 & \Pr\left[\begin{array}{c}
(k-1)\cdot I(X_{1,t};\bar{Y}_{t}|\mathrm{BC\, mode})<R_{1}(D)\leq k\cdot I(X_{1,t};\bar{Y}_{t}|\mathrm{BC\, mode}),\\
R_{2}(D)\leq k\cdot I(X_{1,t};\bar{Y}_{t}|\mathrm{BC\, mode})
\end{array}\Bigg|\, D\right]\nonumber \\
= & \left[F_{S}(q_{k}^{\mathrm{UB}}(D))-F_{S}(q_{k}^{\mathrm{LB}}(D))\right]^{+},\label{eq:Appendix I eq 7}
\end{align}
with $q_{k}^{\mathrm{UB}}(D)$ and $q_{k}^{\mathrm{LB}}(D)$ defined
in Lemma \ref{lem:p2 dec LCSIT}. As a result, we obtain (\ref{eq:p2 dec local CSIT})
by plugging (\ref{eq:Appendix I eq 6}) and (\ref{eq:Appendix I eq 7})
into (\ref{eq:Appendix I eq 5}).

\section{Proof of Lemmas \ref{lem:p2 out adaptive compression} and \ref{lem:p2 dec adaptive compression}}

In this appendix, we derive the results in Lemmas \ref{lem:p2 out adaptive compression}-
\ref{lem:p2 dec adaptive compression} with adaptive compression.

\subsection{Proof of Lemma \ref{lem:p2 out adaptive compression}}\label{appendix:p2 out adaptive compression}

If we assume the adaptive compression described in Sec. \ref{sec:Adaptive Compression},
the probability $p_{\mathrm{out}}^{2}(k)$ is calculated as (\ref{eq:Appendix I eq 4})
with the mutual information $I(X_{2,t};\bar{Y}_{t}|X_{1,t},\mathrm{SL\, mode})$
changed from (\ref{eq:mutual information X2 SL}) to
\begin{equation}
I(X_{2,t};\bar{Y}_{t}|X_{1,t},\mathrm{SL\, mode})=f_{\mathrm{I}}(P,0,\hat{a}_{D},S,D).\label{eq:Appendix II eq 1}
\end{equation}
The only difference from (\ref{eq:mutual information X2 SL}) is the
improved compression gain $\hat{a}_{D}$. If we calculate (\ref{eq:Appendix I eq 4})
with (\ref{eq:Appendix II eq 1}), we immediately obtain the result
in Lemma \ref{lem:p2 out adaptive compression}.

\subsection{Proof of Lemma \ref{lem:p2 dec adaptive compression}}\label{appendix:p2 dec adaptive compression}

With the adaptive compression, the probability $p_{\mathrm{dec}}^{2}(k)$
is given as (\ref{eq:Appendix I eq 5}), and similar to the previous
subsection, the only difference is that the mutual information $I(X_{2,t};\bar{Y}_{t}|X_{1,t},\mathrm{SL\, mode})$
is computed as (\ref{eq:Appendix II eq 1}) with the compression gain
$\hat{a}_{D}$. Then, we can obtain the result in Lemma \ref{lem:p2 dec adaptive compression}.

\section{Proof of Lemmas \ref{lem:p1 out 2 STSC}-\ref{lem:p2 dec 1 STSC}}\label{Appendix:proof for STSC}

In this appendix, we avoid repetition by focusing on the proof of
(\ref{eq:p1 out 2 STSC}) in Lemma \ref{lem:p1 out 2 STSC} since
the proof for Lemmas \ref{lem:p2 out 2 STSC}-\ref{lem:p2 dec 1 STSC}
follows similarly. With STSC, the probability $p_{\mathrm{out}}^{1}(2)$
is given as
\begin{align}
 & p_{\mathrm{out}}^{1}(2)\nonumber \\
= & \Pr\left[I(X_{1,1};\bar{Y}_{1}|\mathrm{BC\, mode})+I(X_{1,2};\bar{Y}_{2}|\mathrm{BC\, mode})<R_{1}\right]\nonumber \\
= & \mathbb{E}_{D_{1},D_{2},S_{1}}\left[\Pr\left[\begin{array}{c}
f_{\mathrm{I}}(\alpha(D)P,\bar{\alpha}(D)P,a_{D_{1}},S_{1},D_{1})+\\
f_{\mathrm{I}}(\alpha(D)P,\bar{\alpha}(D)P,a_{D_{2}},S_{2},D_{2})<R_{1}(D)
\end{array}\Bigg|D_{1},D_{2},S_{1}\right]\right].\label{eq:Appendix III eq 1}
\end{align}
If we express the conditional probability inside the expectation in
(\ref{eq:Appendix III eq 1}) with respect to the CDF of the fading
coefficient $S_{2}$, we get Eq (\ref{eq:p1 out 2 STSC}) in Lemma
\ref{lem:p1 out 2 STSC}.

\end{document}